\documentclass[journal]{IEEEtran}

\usepackage[ruled]{algorithm2e}

\usepackage{latexsym}
\usepackage{algorithmic}
\usepackage{amsmath}
\usepackage{graphicx}
\usepackage{amsfonts}
\usepackage{epsfig}
\usepackage{longtable}
\usepackage{supertabular}
\usepackage{color}
\usepackage{array}
\usepackage{multirow}
\usepackage{booktabs}

\usepackage{setspace}
\usepackage{verbatim}
\usepackage{times}
\usepackage{atbeginend}

\usepackage{multirow}
\usepackage{booktabs}
\usepackage{longtable}
\usepackage{supertabular}

\newcommand{\eat}[1]{}

\SetAlFnt{\small}
\SetAlCapFnt{\small}
\SetAlCapNameFnt{\small}
\SetAlCapHSkip{0pt}

\begin{document}

\title{A Case for Considering Energy Consumption in Software Development and Architecture Decisions}

\author{Kaushik~Dutta
        and~Debra~VanderMeer
\thanks{Kaushik Dutta is with the University of South Florida, e-mail: duttak@usf.edu}
\thanks{Debra VanderMeer is with the Florida International University, email:vanderd@fiu.edu.}}

\maketitle

\begin{abstract}
IT power usage is a significant concern.  Data center energy consumption is estimated to account for 1\% to 1.5\% of all energy consumption worldwide.  Hardware designers, data center designers, and other members of the IT community have been working to improve energy efficiency across many parts of the IT infrastructure; however, little attention has been paid to the energy efficiency of software components.  Indeed, energy efficiency is currently not a common performance criteria for software.  In this work, we attempt to quantify the potential for gains in energy efficiency in software, based on a set of examples drawn from common, everyday decisions made by software developers and enterprise architects.  Our results show that there is potential for significant energy savings through energy-conscious choices at software development and selection time.   
\end{abstract}

\begin{IEEEkeywords}
Energy consumption, Energy efficiency, Energy usage in IT/IS, Software Applications
\end{IEEEkeywords}

\IEEEpeerreviewmaketitle

\begin{spacing}{1.5}
\section{Introduction}
\label{intro}

Recently, the research community has called for a broad research focus on improving IT energy efficiency.   There is good reason for these concerns.  Koomey~\cite{k:08} estimates that power usage by data centers worldwide ranged from 1.1\%  to 1.5\% of all worldwide power consumption in 2010 (roughly 332 billion kWh~\cite{cia_worldConsume:12}).  When converted to CO$_2$ output from electricity generation~\cite{epa_conversion:12}, data center power usage alone generates a volume of greenhouse gases equivalent to 45 million additional vehicles on the road.  To put this number into perspective, that amounts to more than one additional vehicle on the road for every  resident of the U.S. state of California~\cite{census:12}.  

These statistics consider only a portion of computers in use.  In 2008, Gartner estimated that the number of personal computers in use worldwide had passed the one billion mark~\cite{gartnerPC:08}.  Apple reports sales of 75 million iOS (iPhone and iPad) devices in the final quarter of 2012 alone~\cite{apple:12}.  These devices share a general architecture in common with servers in data centers.  Each has a CPU, memory, and persistent storage capable of generic computation.  In addition, each is loaded with a set of software -- an operating system and a set of user-selected applications.  While servers, PCs, and mobile devices may be put to uses ranging from complex financial calculations to high-definition video rendering, each device is a computer that plugs into a socket and draws power and incurs a cost in terms of energy usage.  
  
The issue of energy efficiency is receiving substantial attention along multiple fronts, from data center design to hardware design.  For example, data center designers have worked to improve the infrastructure within data centers to improve the efficiency of cooling systems and power delivery~\cite{openCompute:11,google:11}, while processor designers have proposed chip designs that adjust power use based on workloads~\cite{ctm:00}, and memory designers have improved energy usage by employing multi-level caches (L1 and L2 cache) for frequently accessed datasets in memory~\cite{kgm:97}.

All these solutions impact the IT infrastructure, but do not consider the workload subjected to it, i.e., they do not address the scope of the workload generated by the software that runs on it.  In the context of software, the constructed logic determines the energy usage associated with an application.  These characteristics are determined by those who actually build the software, i.e., the developers and architects.  For example, enterprise architects make broad decisions regarding the choice of operating system and programming language, as well as whether virtualization will be used.  Programmers make thousands of small decisions regarding data types and logic steps.  Each of these choices, both large and small, has an incremental impact on the energy consumption patterns of the final application running in production.  

While the embedded systems software community has for some time held energy efficiency as a standard consideration for development performance, the general software community does not typically consider energy consumption.  Architects and developers make system design choices based on a variety of criteria.  The guideline defined by the ISO international standard ``Systems and software engineering — Architecture description"~p\cite{iso42010:12} outlines a set of forty-eight ``System Considerations" that system designers might consider at build time, including cost, flexibility, affordability, security, and regulatory compliance (among many others).  Energy efficiency is not listed among these system considerations.  In this context, how can we hope to impact energy efficiency in software, if the question of software energy consumption is not even on the table?

 Watson et al.~\cite{wbc:10}[p. 24] ``see the problem as a lack of information to enable and motivate economic and behaviorally driven solutions."~~Murugesan~\cite{m:08}[p. 25] brings the call to a more personal level,  calling on all members of the IT community to focus on what each of us  ``can do individually and collectively to create a sustainable environment."  In the context of software, these individuals are those who design and develop software and software solutions.  Enabling energy-aware decision-making for software, however, is a significant challenge because the information needed to change decsion-making in light of energy consumption is unavailable.   While is is possible to measure overall current drawn by a computer with a watt-meter, it is difficult to map that usage to particular software logic in a program of any complexity bacause modern computer hardware is designed to smooth power flow across time~\cite{sahin2012initial}.   Further, even if it becomes possible to instrument hardware and software to achieve such a mapping, what is a developer to do with such a metric, i.e., how does a programmer change a coding decision based on a watt-meter reading?  A more helpful set of information might present decision-makers with a set of relative energy impacts for common coding patterns.

Our aim in this work is to demonstrate that the effects of software decisions, both large and small in scope, can have a significant impact on energy consumption, with the goal of introducing a new stream of research focused on.  Specifically, in this work, we make the following contributions. 

\begin{itemize}

\item 

We first consider decision-making at the developer level, looking at a set of the basic building blocks of an application.  Here, we demonstrate that even small, seemingly innocuous choices can be associated with substantial differences in energy usage, differences that can add up to significant disparities in energy consumption over the course of an application's lifetime.

\item

We then consider more complex applications, representative of the types of software running in organizations, and consider the energy consumption impacts of a set of decisions commonly made by an enterprise architect, and show that common architectural configurations are associated with significantly different energy consumption profiles.   

\item

Finally, we discuss the potential benefits of taking energy efficiency into account across the software stack, and argue that energy efficiency is a useful performance concern in its own right.  

\end{itemize}

The remainder of this paper is organized as follows.  In Section~\ref{related}, we discuss our work in the context of related work.  In Section~\ref{method}, we describe the methodology we followed in our experiments.  In Sections~\ref{developer} and~\ref{architect}, we discuss the energy impacts of common developer decisions and common decisions made by architects, respectively.  In Section~\ref{discussion}, we discuss the potential financial and environmental benefits associated with more thoughtful energy-related software choices.  Finally, we conclude in Section~\ref{conclusion}. 

\section{Related Work}
\label{related}
Energy usage IS/IT has emerged as an important area of research area for the technology community, with broad calls for more environmentally-friendly information technology from both industrially-oriented researchers~\cite{Gartner_greenIT:07} and public policy analysts~\cite{ruth2009green}. Indeed, a broad set of researchers have begun to study information technology in the context of environmental sustainability. In this section, we survey work in this area, and relate the broad streams of research to our work here. 

At a high level, work focused on the use of information systems to support environmental decision-making is related. This is a broad stream of research, primarily led by researchers in geography and geoinformatics. Some examples of this type of work include Tsou's~\cite{tsou2004integrated} work in developing integrated mobile GIS tools for environmental monitoring and management field work, Pundt and Bishr~\cite{pundt2002domain}'s research on developing ontologies to facilitate sharing environmental field data, Ceccato et al.'s~\cite{ceccato2005application} work on remote sensing systems to monitor the risk of malaria spread in vulnerable areas in the developing world, and Kingston's~\cite{kingston2000web} work on developing public-participation systems to support inclusive environmental decision-making. While related, this work is complementary to our work, i.e., we focus the energy efficiency of the software systems themselves, rather than using the systems for environmental monitoring/management purposes.

A number of studies have looked at using hardware more efficiently, and taking advantage of recent developments in hardware for energy-awareness. New scheduling algorithms have been proposed for resource-conscious CPU scheduling for multi-core processors~\cite{merkel2010resource} and energy efficiency for CPU-constrained tasks with real-time deadlines~\cite{wu2004cpu}, as well as to take advantage of recent Dynamic Voltage Scaling features in CPUs to compile code with hints for CPU step-downs~\cite{hsu2003design} (which is particuarly important in the area of low-power high-performance computing), and to balance loads across servers in a cluster for energy efficiency~\cite{rusu2006energy}. Other studies propose low-level memory management techniques~\cite{tolentino2009memory,balasubramonian2000memory,kin2000filtering} for energy efficiency. Energy-efficient disk management techniques have been proposed, including an energy-optimized RAID architecture~\cite{mao2008graid}. Energy-efficient network management techniques have been proposed for wireless networks~\cite{uysal2002energy} and ad hoc sensor networks~\cite{younis2004heed}.  Finally,~\cite{rivoire2007joulesort} propose an energy efficiency benchmark designed to serve as a means of comparing different hardware platforms. This stream of work is also complementary to ours; while this work works to improve hardware-related energy efficiency, our focus is on increasing software energy efficiency. 

In the realm of software energy efficiency, the most prominent stream of research concerns energy efficiency for embedded systems.  These systems are typically designed for mobile or remote applications, where battery life is a significant concern -- lower energy consumption translates to longer device life.  Further, the software in an embedded system is often etched in a chip, such that software updates are not possible; once software design choices are made, there is no easy way to modify a high-energy-consumption design decision with a lower-consuming one.   The literature contains a wide variety of work in this area.  We provide a brief survey to illustrate some of the types of issues addressed by the research in this domain.  Lekatasas et al.~\cite{lekatsas2000code} proposes a novel method for incorporating instruction code compression to reduce power usage in embedded systems.  Shiue and Chakrabarti~\cite{shiue1999memory} describe a method for choosing a power-optimal memory configuration for an embedded system, given the system's specific characteristics.  Peymandoust et al.~\cite{peymandoust2002low} proposes an algebraic method for optimizing complex instructions (which are energy-intensive) for embedded software.  Choi and Chatterjee~\cite{choi2001efficient} describe a method for optimizing instruction ordering in embedded software to reduce energy requirements.  Farinelli et al.~\cite{farinelli2008decentralised} propose a method for coordinating among decentralized embedded devices (a common requirement in sensor systems) that optimizes for reduced energy usage.  A common characteristic of most work in this area is that it addresses architectural issues in the embedded system; in other words, it addresses issues surrounding how the software logic is handled after the logic has been defined, rather than looking at the energy implications within the logic itself.  Other studies consider how energy simulation for embedded systems can improve energy efficiency~\cite{simunic2001energy}, as well as  energy-efficient routes in ad hoc networks, energy-efficient placement of distributed computation, and flexible RPC/name binding~\cite{vahdat2000every}.  

Aside from the work in embedded systems, a search of the literature revealed little research directly addressing energy efficiency in software, specifically in common application software.  A further study~\cite{capra2009green} proposes a framework to describe theoretical bounds on software energy based on the concept of thermodynamic depth.  While this study is valuable, the hypotheses presented remain untested in any practical way.  To summarize, virtually all work that considers energy consumption in software is either limited to a narrow type of software application, or remains in the realm of theory.  In contrast, in our work, we aim for a practical contribution: to show experimentally that common, everyday decisions regarding software applications can significantly impact software energy consumption. 

Finally, there are also tools available to help measure system energy usage, including support for comparative analysis and measurement.  To support comparative analysis, for example, the Transaction Processing Performance council has issued the TPC-Energy Specification~\cite{tpc-energy:10}.  This specification provides a common software workload and methodology for comparing energy consumption across different hardware platforms.  In terms of measurement support, Intel provides an Energy Checker SDK~\cite{intel-energy:10} to support energy consumption measurement.  Specifically, the SDK provides counters that software developers can embed into code to monitor the number and timing of invocations of different portions of code.  When coupled with data gathered by an external power meter (which measures actual consumption), developers can build metrics of applicaiton energy usage.  For example, a developer working on email server software might insert counters to track counts of emails sent or volume of data transferred in emails, and use these counters to devleop aggregated measures of work (emails/data transferred) per unit of consumption. 

The IS organizational community has produced a number of studies looking at environmental sustainability for businesses.  Melville~\cite{melville2010information} and Elliot~\cite{elliot2011transdisciplinary} provide good overviews of the area in the form of meta-studies. Jenkin et. al~\cite{jenkin2011agenda} and Dedrick~\cite{dedrick2010green} describe research frameworks for green IT and green IS respectively. Theoretical studies describe firms' preparedness to adopt energy-aware IS/IT initiatives in terms of organizational readiness~\cite{molla2009and}, institutional forces that shape environmental concerns~\cite{butler2009environmental}, organizational motivation~\cite{chen2008information}, managerial attitudes toward green IT~\cite{sarkar2009managerial}, as well as potential to undertake energy-related IT initiatives via virtualization~\cite{bose2011integrative}. Predictive studies have used empirical~\cite{schmidt2010predictors} and simulation~\cite{hilty2006relevance} methods to forecast organizational green IT adoption. A further set of studies~\cite{dao2011green},~\cite{darnall2008environmental},~\cite{el2006environmental} consider how organizations can operationalize their environmental management processes and systems with a focus on sustainability.  This set of work provides a framework to motivate our own work here, in that this theoretical work demonstrates the need for energy efficiency, while our work seeks to begin to address that need.  

\section{Methodology}
\label{method}

\begin{figure*}
	\begin{minipage}[b]{0.5\linewidth}
	\centering
	\epsfig{file=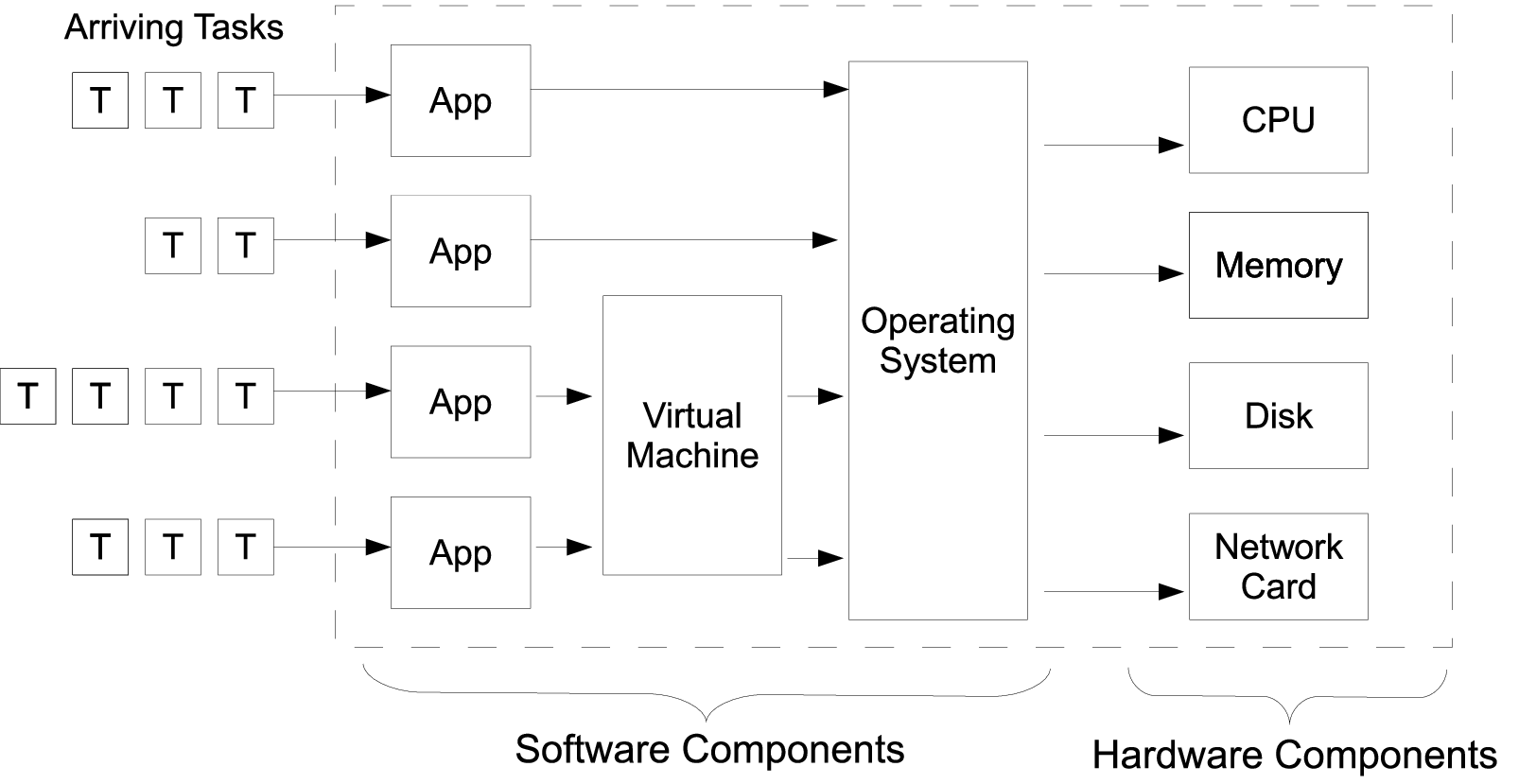,height=1.5in}
	\caption{Energy Model for a Hardware/Software Stack}
	\label{fig:energyModel}
	\end{minipage}
	\begin{minipage}[b]{0.5\linewidth}
	\centering
	\epsfig{file=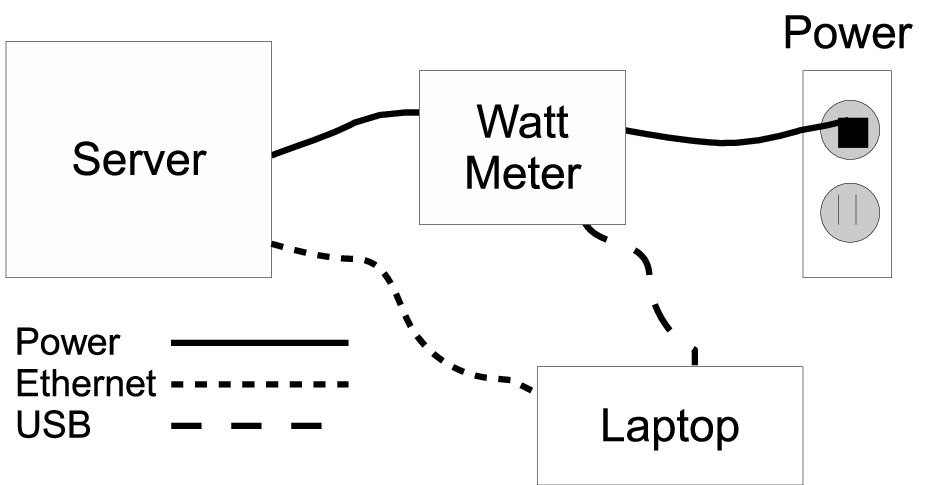,height=1.5in}
	\caption{Experimental Architecture}
	\label{fig:expArch}
	\end{minipage}
\end{figure*}

To commence our exploration of software energy usage, we begin with a look at a typical hardware/software stack, as depicted in Figure~\ref{fig:energyModel}.  Fundamentally, the only components that actually consume electricity are the hardware components, where the CPU navigates a set of physical logic gates to perform calculations and modify data in its registers, the RAM components change the contents of their memory cells, the disk head magnetically manipulates sectors on the disk, and the network interface card sends and receives data packets over a physical cable connection or via a wireless antenna.  

All of the software components, including all applications, any application servers and virtual machines, and the operating system, are simply compiled executable files stored on a computer's disk.  At runtime, the physical hardware components load these files from the disk into RAM for execution on the CPU.  The software itself does not consume energy; rather, the logic contained within the compiled code in the software files drives the hardware components, which actually consume power.  Thus, for the purposes of this work, when we use the phrases ``software energy consumption" or ``software power usage," we are referring to the power consumption of the hardware components driven by the execution of the logic encoded in the software components.

Our survey of software energy consumption consists of a set of experiments looking at the power usage impacts of choosing among a set of typical options for a set of decisions common to the roles of developer and enterprise architect. Figure~\ref{fig:expArch} depicts the basic experimental architecture for all the experiments we describe in this work. A server forms the main experimental platform where we run different types of logic, and observe the energy consumption impacts.  To measure power consumption on the server, we inserted a Watt-meter component, specifically a ``Watts up? PRO'' power meter~\cite{wattsup:11} between the server and the power outlet.  We configured the Watt-meter to measure fine-grained power consumption statistics and store them on the laptop (via USB cable).


We begin our investigation with a set of experiments that consider choices that developers make on a daily basis.  We look at some of the basic logic building blocks of any application, including the creation of objects, the variety of numeric data types, common string representations, and a set of different libraries for the same task: sorting.  For each of these dimensions, a developer makes a selection among several available choices.  In these experiments, we demonstrate the energy implications of each option for each dimension.  To isolate the effects of each option, we run each experiment as a set of sequential operations in a single thread, with no other processes running on the server (other than those required for the operating system).  We present the results of these experiments in Section~\ref{developer}, and discuss the implications.

Next, we consider a more complex software scenario, and look at the energy consumption implications associated with a common decision made by enterprise architects, namely the choice of which operating system platform to use for a system.  Given the prevalence of virtualization technologies, we also consider a set of common operating system configurations for hosted OSs. In these experiments, we configured the server in our experimental architecture (in Figure~\ref{fig:expArch}) with an application server and a benchmark web application.  We configured the laptop with the JMeter load testing software to generate a varied workload on the web application.  We then run a set of experiments that gather both energy consumption and response time data as the web application is subjected to a series of increasing workload scenarios.  We vary the workload by varying the number of simultaneous JMeter users accessing the web site's functionality.  In each experimental case, we vary the the choice of operating system and virtualization configuration.  We present the results of these experiments in Section~\ref{architect}, and consider the associated implications.

\section{Energy Impacts of a Developer's Decisions}
\label{developer}
In this section, we demonstrate the differences in energy consumption associated with a set of decisions that developers make on an everyday basis.  We consider the following scenarios:  (1) object creation and reuse; (2) choosing a numeric data type; (3) choosing a character string representation; and (4) choosing a sorting method, an example of a common developer decision with multiple choices available across a set of standard libraries.   

We developed a set of programs designed to test these software design choices in isolation to observe the energy consumption impacts of each option.  We coded each experiment in Java 1.6, and ran the experiments.  During each experiment, we recorded the server's energy consumption.  With no user applications running, the server (a Dell Optiplex computer with dual-core 2.8 GHz CPU and 4 GB RAM, running Windows Server 2008) consumes 84.63 watts on average.  In our experimental results, we reported the total energy consumption of the server for exclusively running each experiment. In our results, we report the total energy consumption of the server, including the baseline server energy consumption, rather than reporting the difference above the baseline for each experimental case. 

In these experiments, operations run on randomly-generated values.  Numeric values were generated using the Java Random class based on a uniform distribution as follows: the {\tt float} and {\tt double} range is 0.0 to 0.1, while the {\tt int} range is $-2^{31}$ to $2^{31}$.  Characters were randomly selected from the ASCII character set. Our results reported here do not include the energy cost of generating values.  In each experiment, we first pre-create the appropriate number of values required, and then perform the operations of interest.  We report energy consumption only for the experimental period when the operations of interest are running (the power meter's time is synchronized to the laptop's time via USB connection).  For each experimental case, we run the experiment 10 times, and report the average energy consumption in our results. 

We now move on to describe our experimental results. 

\subsection{Comparing Energy Consumption for Object Creation and Reuse}

In this experiment, we explore the energy consumption effects of reusing an instantiated object, rather than creating multiple instantiations.  We created a Java class called {\tt Simple} with a single {\tt int} data member, a simple unparameterized constructor containing no logic, and a {\tt set} method for the data member.  We created two Java console applications, each containing a single loop.  In the object creation case, a {\tt Simple} object is instantiated within the loop, and its data member is set to a random integer value.  In the object reuse case, the {\tt Simple} object is instantiated prior to the loop; inside the loop, its data member is set to a randomly-generated integer.

\begin{table*}%
\begin{footnotesize}
\begin{center}
\caption{Energy Consumption for Object Creation and Object Reuse\label{tab:object_creation}}{%
\begin{tabular}{rrrrr}
    \addlinespace
    \toprule
    \textbf{Task} & \textbf{Number of} & \textbf{Duration} & \textbf{Joules} & \textbf{Average \% Energy Savings } \\
   & \textbf{Objects} & \textbf{(ms)} & \textbf{(Watt-Sec)} & \textbf{in Object Reuse} \\
   &  &  &  & \textbf{Compared to Object Creation} \\
    \midrule
    \multicolumn{1}{l}{\multirow{4}[0]{*}{Object Creation }} & 100,000 & 9     & 0.7497 &  \\
    \multicolumn{1}{l}{} & 1,000,000 & 30    & 2.517 &  \\
    \multicolumn{1}{l}{} & 10,000,000 & 226   & 18.9614 &  \\
    \multicolumn{1}{l}{} & 100,000,000 & 20582 & 2410.1522 &  \\ \midrule
    \multicolumn{1}{l}{\multirow{4}[0]{*}{Object Reuse}} & 100,000 & 6     & 0.4998 & 33.33 \\
    \multicolumn{1}{l}{} & 1,000,000 & 10    & 0.839 & 66.67 \\
    \multicolumn{1}{l}{} & 10,000,000 & 40    & 3.356 & 82.30 \\
    \multicolumn{1}{l}{} & 100,000,000 & 993   & 118.5642 & 95.08 \\
    \bottomrule
\end{tabular}}
\end{center}
\end{footnotesize}
\end{table*}%

We report the average energy consumption for the creation and reuse cases in Table~\ref{tab:object_creation}.  In each experiment, we ran  $10^I$ iterations of the loop, where $I = 5, 6, 7, 8$.  Our results show that there is a significant energy consumption difference between the creation and reuse cases, with the reuse case showing 95\% energy savings over the creation case for $I=8$.  Clearly, there is a substantial difference between reusing instantiated objects and creating them. 

\subsection{Comparing Energy Consumption across Numeric Data Types}

In this experiment, we consider energy consumption for mathematical operations across the four commonly-used data types in application programs: \texttt{int}, \texttt{float}, \texttt{double} and \texttt{boolean}. For each data type, we perform mathematic operations across $M$ pairs of data values. For \texttt{int}, \texttt{float} and \texttt{double} values, the operation is selected from the following operations: addition, subtraction, multiplication and division. For \texttt{boolean}, the operation is selected from the following operations: AND, OR, NOT and XOR. 

We ran the experiment for two values of $M$, 100 million and 1 billion. We report the average time to run and energy consumption in Table~\ref{tab:datatype}. 

\begin{table*}%
\begin{center}
\begin{footnotesize}
\caption{Comparing Energy Consumption across Numeric Data Types\label{tab:datatype}}{%
\begin{tabular}{rrrrrr}
    \addlinespace
    \toprule
    \textbf{Number of}  & \textbf{Data}  & \textbf{Time of Computation}  & \textbf{Average Power} & \textbf{Energy Consumption} & \textbf{\% Energy Savings} \\
    \textbf{Operations}   &   \textbf{Type} & \textbf{(milliseconds)} & \textbf{(Watt)}  & \textbf{(Joules)}  & \textbf{in Integer} \\ 
    \midrule \midrule
    \multirow{4}[0]{*}{100,000,000} & \texttt{int} & 6074  & 85.7  & 520.5418 & 0 \\
          & \texttt{float} & 6473  & 91.34 & 591.24382 & 11.96 \\
          & \texttt{double} & 10426 & 91.77 & 956.79402 & 45.60 \\
          & \texttt{boolean} & 6066  & 91.56 & 555.40296 & 6.28 \\ \midrule
    \multirow{4}[0]{*}{1,000,000,000} & \texttt{int} & 60610 & 85.81 & 5200.9441 & 0 \\
          & \texttt{float} & 64615 & 92.29 & 5963.31835 & 12.78 \\
          & \texttt{double} & 104157 & 92.37 & 9620.98209 & 45.94 \\
          & \texttt{boolean} & 60670 & 92.33 & 5601.6611 & 7.15 \\
    \bottomrule
\end{tabular}}
\end{footnotesize}
\end{center}
\end{table*}%

Table~\ref{tab:datatype} shows that there is considerable difference in energy consumption associated with the data types. Computations involving \texttt{int} require 12-13\% less energy than \texttt{float}. Computations for \texttt{int} require 46\% less energy than \texttt{double} data types.  

Surprisingly, the computational overhead of \texttt{int} operations is 6.3\% less than that of \texttt{boolean} operations, even though \texttt{boolean} is logically a one bit data type.  Intuitively, it would seem more logical that comparing two single-bit values should require less energy than mathematical operations across 32-bit values.  However this ignores that the fact that the {\tt boolean} data type is more complex than it seems to be at first glance. This is because it is emulated by a 1 byte data type, which needs to ensure that only the lowest bit is set to 0 or 1, (false or true), and that the rest of the bits are set to always 0. This extra logic in the {\tt boolean} data type is associated with additional overhead, which is responsible for the additional energy consumption in the {\tt boolean} case as compared to the {\tt int} scenario. 

In application programs, a \texttt{boolean} data type can be replaced by an \texttt{int} with \{0, 1\} value.  This is a common practice in many programming scenarios -- the C language does not even have a standard boolean data type.  

Our results show that considerable energy consumption could be saved when the \texttt{double} data type is used and an \texttt{int} or \texttt{float} would suffice for the purpose, as well as when the \texttt{boolean} data type is used and the same functionality can be achieved with an \texttt{int}.   

\subsection{Comparing Energy Consumption across String Representations}

In this experiment, we consider the energy consumption impacts of two different string representations:  {\tt String} and {\tt StringBuffer}.  In Java, a {\tt String} is immutable; changing it requires creating a new {\tt String} object.  In contrast, {\tt StringBuffer} allows changes to a sequence of characters.   For these experiments, we created two Java console applications, each containing a single loop.  In the string case, a {\tt String} object is instantiated prior to the loop, and a single randomly-generated character is appended to the {\tt String} in each loop iteration.  In the string buffer case, a {\tt StringBuffer} object is instantiated prior to the loop, and a single randomly-generated character is appended to the {\tt StringBuffer} in each loop iteration.  We ran this experiment  for 10,000, 100,000 and 1 million loop iterations.  Table~\ref{tab:string} shows the results of these experiments. 

\begin{table*}%
\begin{center}
\begin{footnotesize}
\caption{Energy Consumption across String Representations\label{tab:string}}{%
\begin{tabular}{rrrrr}
    \addlinespace
    \toprule
    \textbf{Task} & \textbf{Number of} & \textbf{Duration} & \textbf{Joules} & \textbf{Average \% Energy Savings} \\
    & \textbf{Objects} & \textbf{(ms)} & \textbf{(Watt-Sec)} & \textbf{in StringBuffer} \\
    & & & & \textbf{Compared to String Creation} \\
    \midrule
    \multicolumn{1}{l}{\multirow{3}[0]{*}{String}} & 10,000 & 159   & 18.921 &  \\
    \multicolumn{1}{l}{} & 100,000 & 7789  & 926.891 &  \\
    \multicolumn{1}{l}{} & 1,000,000 & 1144689 & 136217.991 &  \\ \midrule
    \multicolumn{1}{l}{\multirow{3}[0]{*}{StringBuffer}} & 10,000 & 2     & 0.238 & 98.74 \\
    \multicolumn{1}{l}{} & 100,000 & 8     & 0.952 & 99.89 \\
    \multicolumn{1}{l}{} & 1,000,000 & 23    & 2.737 & 99.99 \\
    \bottomrule
\end{tabular}}
\end{footnotesize}
\end{center}
\end{table*}%

The difference in energy consumption between {\tt String} and {\tt StringBuffer} data types is significant -- 98\% less energy is consumed in the string buffer case across 10,000 append operations, as compared to energy consumption in the string case.  While the {\tt StringBuffer} data type is slightly more complex work with in code, the energy consumption overheads associated with the {\tt String} data type more than outweigh the implementation convenience the {\tt String} type offers.  

\subsection{Comparing Energy Consumption across Sorting Methods}

In these experiments, we compared the energy consumption impacts of three different types of sorting: (a) {\tt Qsort}, an implementation of the quicksort algorithm~\cite{h:62}, which is known to give on average $O(n~ log~n)$ complexity and is the most widely used sort algorithm in practice; (b) Java's \texttt{Array.sort}, which is a tuned quicksort, adapted from~\cite{arraysort}; and (c) Java's \texttt{Collections.sort}, which is a modified mergesort where the merge is omitted if the highest element in the low sublist is less than the lowest element in the high sublist.

In each experiment, we apply each of the three sort options to sort $N$ integers. We ran the experiment for three values of $N$ (10 million, 50 million, and 100 million).  We measured the total power usage in the computer during the sort operation, and report our results in Table~\ref{tab:sort}.

\begin{table*}%
\begin{center}
\begin{footnotesize}
\caption{Energy Consumption across Sorting Methods\label{tab:sort}}{%
\begin{tabular}{rrrrrr}
    \addlinespace
    \toprule
    \textbf{Number of} &  \textbf{Algorithm}  & \textbf{Time to Run} & \textbf{Average Power}  & \textbf{Energy Consumption} & \textbf{\% Energy Savings}  \\
       \textbf{Integers} &   &  \textbf{(milliseconds)} &  \textbf{(Watt)} & \textbf{(Joules)} & \textbf{compare to \texttt{Qsort}} \\ \midrule \midrule
    \multirow{3}[0]{*}{10,000,000} & \texttt{Qsort} & 1590  & 83.7  & 133.083 &  0\\
          & \texttt{Arrays.sort} & 74    & 83.7  & 6.1938 &  95.35 \\
          & \texttt{Collections.sort} & 1175  & 93.4  & 109.745 &  17.54\\ \midrule
    \multirow{3}[0]{*}{50,000,000} & \texttt{Qsort} & 8635  & 94.036 & 812.0009 & 0\\
          & \texttt{Arrays.sort} & 163   & 97.3  & 15.8599 &  98.05\\
          & \texttt{Collections.sort} & 1498  & 112.85 & 169.0493 &  79.18\\ \midrule
    \multirow{3}[0]{*}{100,000,000} & \texttt{Qsort} & 17799 & 97.8  & 1740.742 &  0\\
          & \texttt{Arrays.sort} & 276   & 98.3  & 27.1308 &  98.44\\
          & \texttt{Collections.sort} & 3044  & 118.68 & 361.2619 & 79.25\\
    \bottomrule
\end{tabular}}
\end{footnotesize}
\end{center}
\end{table*}%

Based on Table~\ref{tab:sort}, we can see that \texttt{Arrays.sort} provides the most efficient performance in terms of energy consumption, showing energy savings in the  range of 95-98\% compared to the generic \texttt{Qsort} algorithm. \texttt{Collections.sort} is more efficient than  \texttt{Qsort}, but less efficient than \texttt{Arrays.sort}.  For lower values of $N$,~\texttt{Collections.sort} consumes 17\% less energy than \texttt{Qsort}; for 50 million or above integers it uses 80\% less energy than \texttt{Qsort}. The energy savings reported here is primarily due to reductions in the running time of the algorithm in the \texttt{Arrays.sort} and \texttt{Collections.sort} cases. 

While the \texttt{Qsort} algorithm is one of the most popular and widely used algorithms in practice, both \texttt{Collections.sort} and \texttt{Arrays.sort} are sort implementations easily available in one of the most widely used development frameworks, the Java Development Kit.  In such a situation, a less-experienced developer could easily select a sort implementation that works functionally, but comes with a higher cost in terms of energy usage. 

\subsection{Summary}

In these experiments, we ran each scenario in a loop over thousands of invocations to show the impact of developer decisions over time.  Software, once compiled, may run 24/7/365, so the choice that a developer makes at build time may have billions or trillions of impacts over time.  Any deviation from the most energy efficient option possible in a given development scenario can result in significant energy impacts at runtime over the course of days, weeks, and years.

\section{Energy Impacts of an Architect's Decisions}
\label{architect}
In this section, we consider the perspective of an enterprise architect, who makes an organization's high-level strategic IT decisions.  One example of such a decision involves choosing a reference architecture for the organization.  In this context, some of the fundamental decisions involve operating systems and virtual machines. Here, we consider the energy impacts of some basic decisions an enterprise architect needs to make:
\begin{itemize}
\item What operating system should we choose in our reference architecture? 
\item Should we choose a virtual machine-(VM)-based architecture? 
\item If we choose a VM-based architecture, what operating system should we use in the VM?
\end{itemize} 

The choices made for each of these questions have long-term implications along several dimensions, including maintenance, licensing, scalability and performance, and operational costs, all of which play a role in the architect's decision. In this section, we demonstrate that these decisions  also have non-trivial consequences in terms of energy consumption.  We validate this claim with a set of experiments, which we describe below. 

\subsection{Experimental Setup}

In these experiments, we follow the same basic experimental architecture described in Section~\ref{method}, where the main server is a Dell Optiplex computer with quad-core 2.8 GHz CPU and 8 GB RAM.  

To consider an application and workload scenario that is closer to real-life than the simple single-function experiments in Section~\ref{developer}, we created an experimental testbed with an application server (Apache Tomcat 2.2 with PhP 5.3 using the PHP/Java Bridge~\cite{php-java:15}), a database (MySQL 5.1), and a benchmark web application.   

For the benchmark web application, we configured the RUBiS~\cite{rubis:13} benchmark application on the server in our experimental testbed.  RUBiS is an auction site benchmark similar to eBay.com, and is widely used to evaluate application servers' performance. It implements the basic functions of an auction site, e.g., browsing, logging in, and selling. We use this benchmark to represent a typical online scenario, where many users interact with the site simultaneously. 

We generated varying workloads on the RUBiS application by emulating multiple simultaneous users using JMeter~\cite{jmeter:13} (a load testing tool).  We configured the laptop (a dual-core 2.8 GHZ CPU and 4 GB RAM machine) in the base experimental platform depicted in Figure~\ref{fig:expArch} with JMeter to generate the request workload.  To avoid network latency overhead, the JMeter computer and the RUBiS server were connected using a local switch in the same network. 

In these experiments, we studied all three types of actions in the RUBiS benchmark (i.e., logging in, selling and browsing) separately, and measured the growth of application response time (using JMeter) and energy consumption of the server (using a ``Watts up? PRO'' power meter~\cite{wattsup:11}) as the RUBiS application is subjected to increasing workloads (simulated by increasing the number of simultaneous JMeter users accessing the target web application).  Each action (Logging in, Selling, and Browsing) was tested with different workload levels, which were realized by a different amount of simultaneous users (5-1000). For each emulated user, JMeter sends a request to the target RUBiS application.  Once the JMeter user receives a response from the application, it waits for a think time between 10 and 50 milliseconds (determined by JMeter's Gaussian Random Timer), and then sends another request with different parameters to the application. We conducted ten fourteen-minute measurement experiments for each experimental case. Typically, a measurement experiment begins with a set up phase that starts a workload generation thread to represent each simulated user. A stabilization phase of two minutes ensures a stable workload generation. The rating period for each experiment lasted ten minutes. Finally, a supplementary phase of two minutes ensures that the workload does not break down abruptly at the end of the rating period. All threads are shut down after the supplementary run. 

For each RUBiS activity, we consider six different O/S and VM configurations, where the application server and benchmark web application were installed on the native O/S if there was no VM configured, or on the VM if one is configured:  Windows-No VM (W); Linux-No VM (L); Windows-Windows VM (WW); Linux-Linux VM (LL); Windows-Linux VM (WL); and Linux-Windows VM (LW). For all Windows cases, the O/S was Window Server 2008; for Linux cases, the O/S was Ubuntu 12.0.  In each case, the experiment was run 10 times, and the average values are reported. At idle condition (when no users are connected to the RUBiS application), the server consumed 65 watts on average. 

\subsection{Experimental Results}

\begin{figure*}
	\begin{minipage}[b]{0.5\linewidth}
	\centering
	\epsfig{file=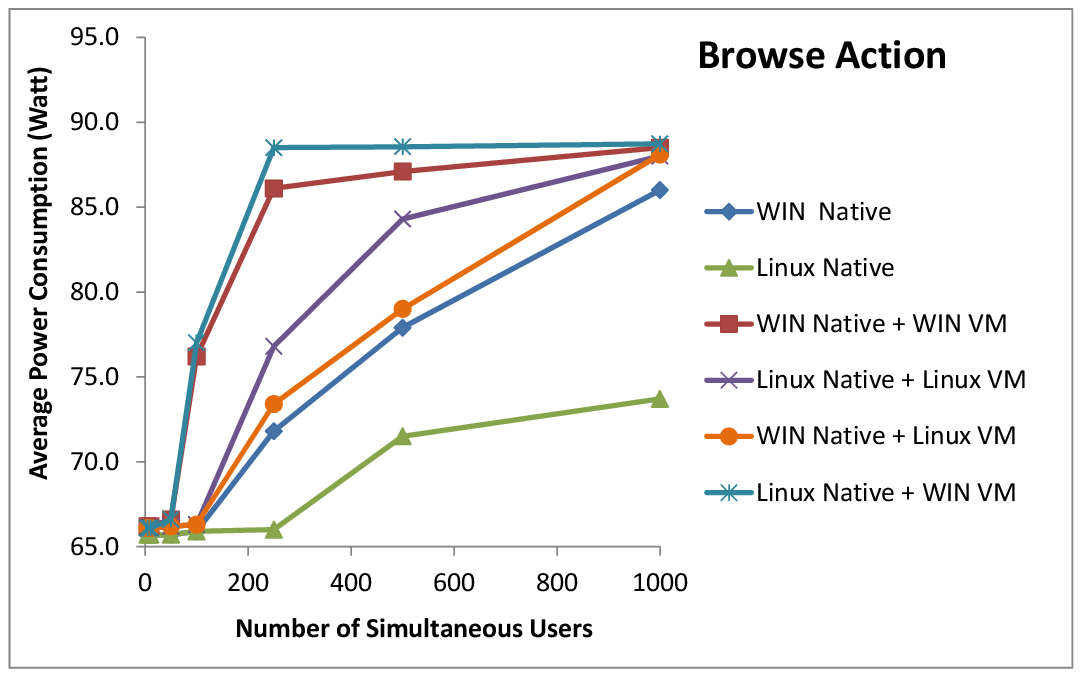,height=2.2in}
	\caption{Average power characteristics for Browse actions across different OSs}
	\label{fig:browse-power}
	\end{minipage}
	\begin{minipage}[b]{0.5\linewidth}
	\centering
	\epsfig{file=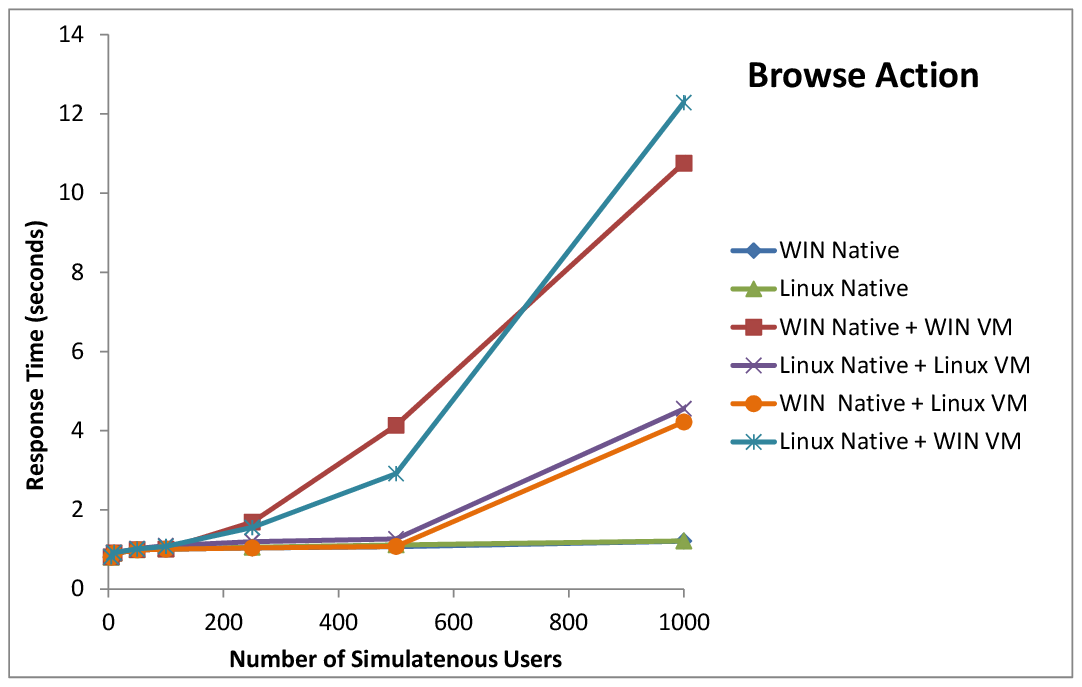,height=2.2in}
	\caption{Response time characteristics for Browse actions across different OSs}
	\label{fig:browse-rt}
	\end{minipage}
\end{figure*}

\begin{figure*}
	\begin{minipage}[b]{0.5\linewidth}
	\centering
	\epsfig{file=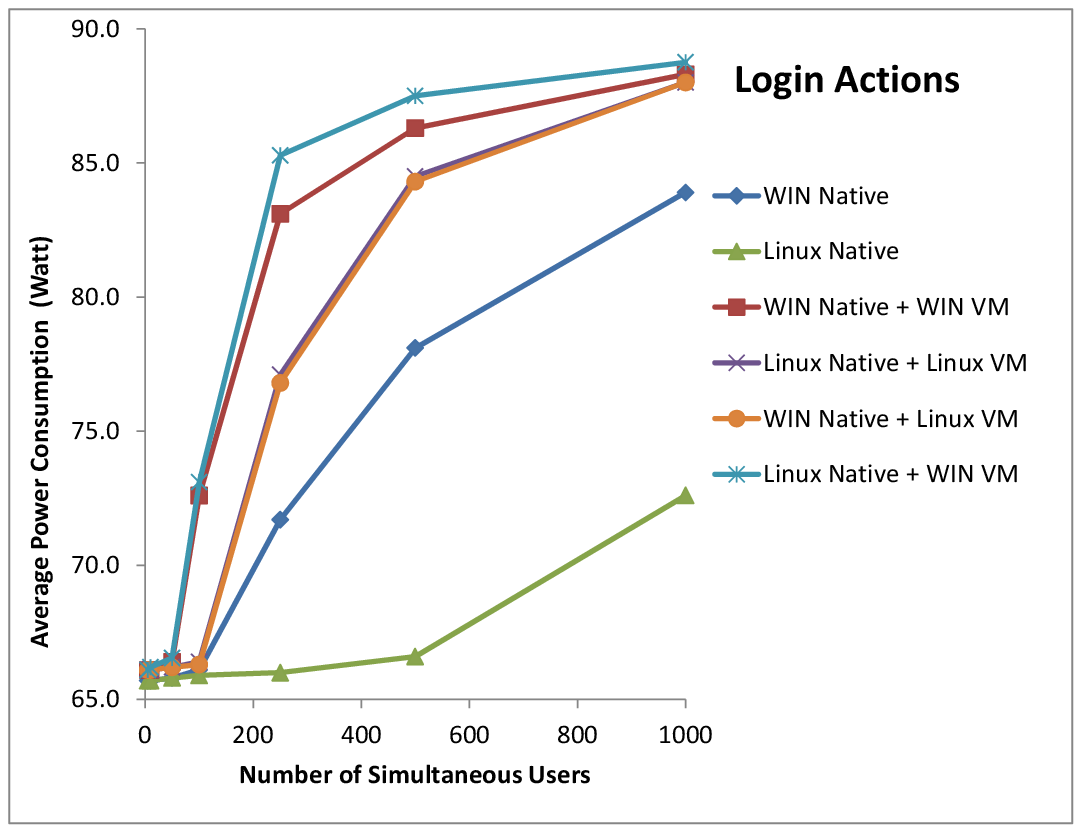,height=2.5in}
	\caption{Average power characteristics for Login actions across different OSs}
	\label{fig:login-power}
	\end{minipage}
	\begin{minipage}[b]{0.5\linewidth}
	\centering
	\epsfig{file=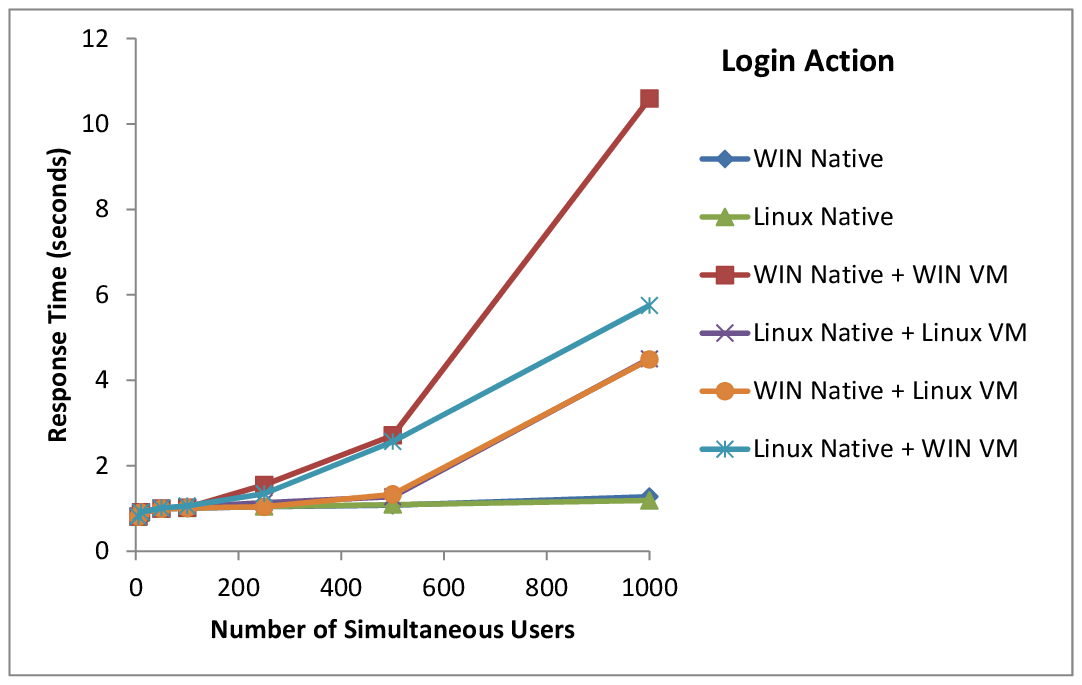,height=2.5in}
	\caption{Response time characteristics for Login actions across different OSs}
	\label{fig:login-rt}
	\end{minipage}
\end{figure*}

\begin{figure*}
	\begin{minipage}[b]{0.5\linewidth}
	\centering
	\epsfig{file=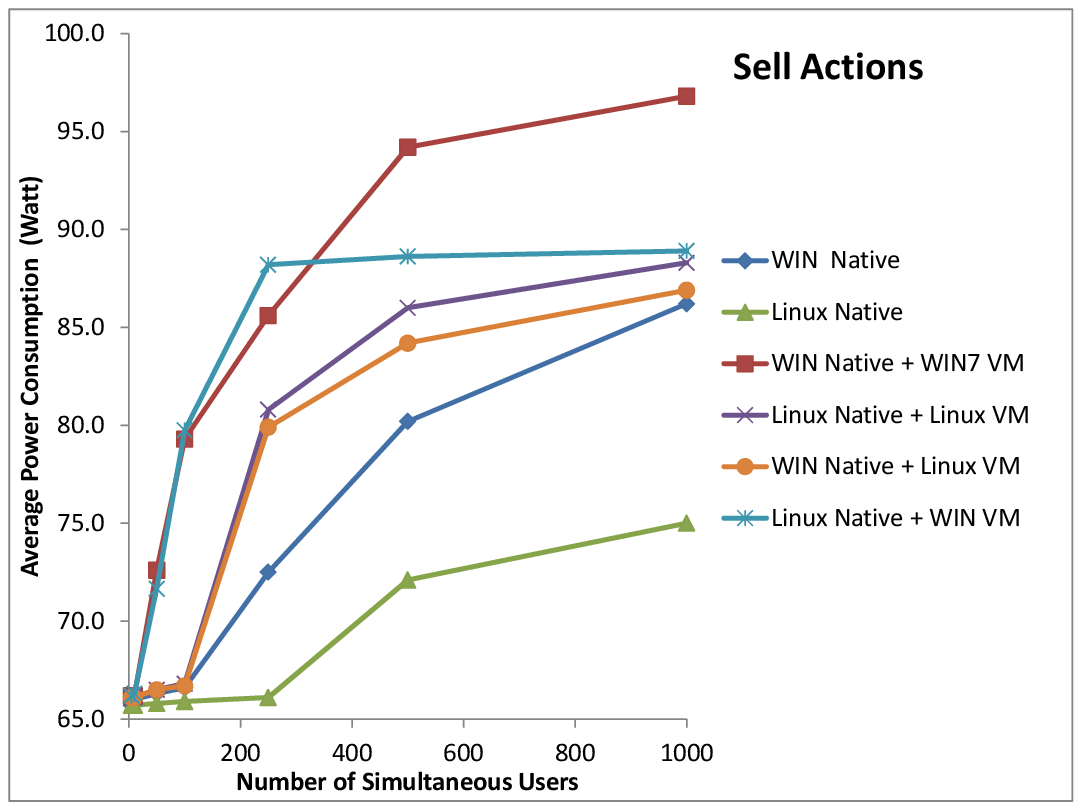,height=2.4in}
	\caption{Average power characteristics for Sell actions across different OSs}
	\label{fig:Sell-power}
	\end{minipage}
	\begin{minipage}[b]{0.5\linewidth}
	\centering
	\epsfig{file=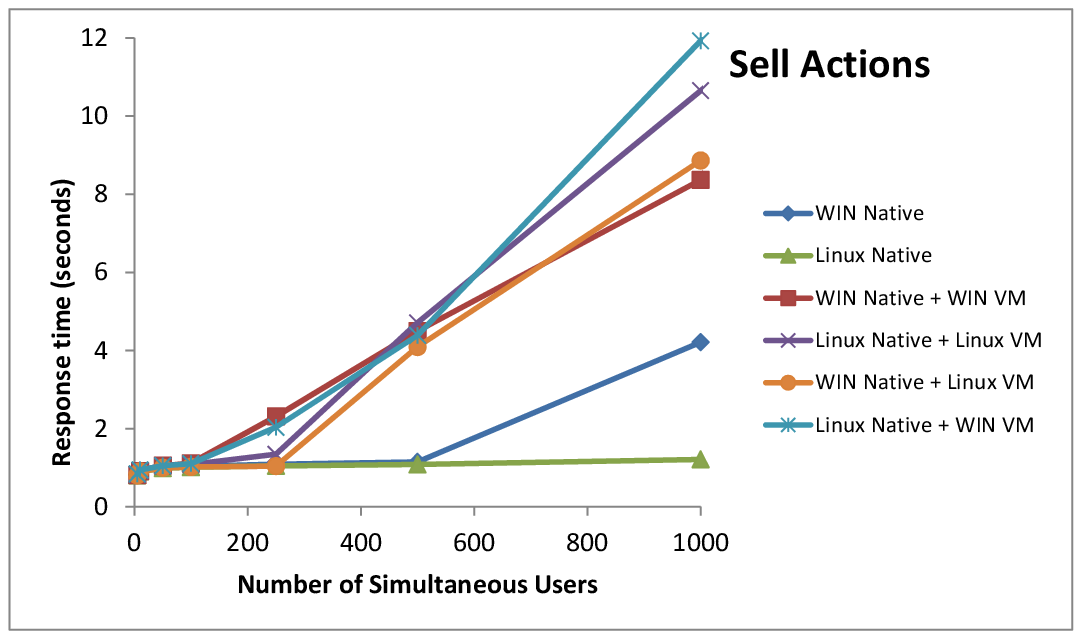,height=2.4in}
	\caption{Response time characteristics for Sell actions across different OSs}
	\label{fig:Sell-rt}
	\end{minipage}
\end{figure*}

\begin{figure*}
	\begin{minipage}[b]{0.5\linewidth}
	\centering
	\epsfig{file=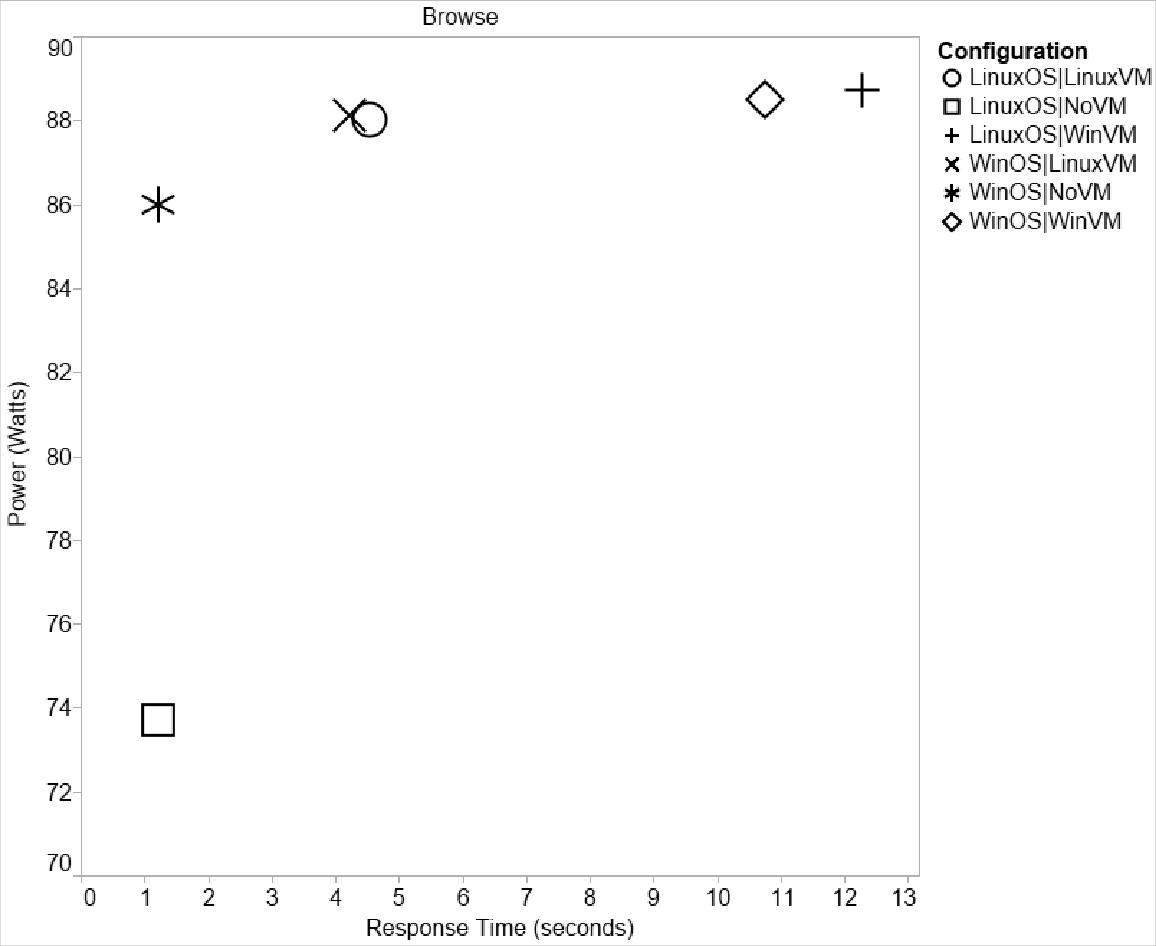,width=2.7in}
	\caption{Response time vs. Energy characteristics for Browse actions}
	\label{fig:Browse_RT_Energy}
	\end{minipage}
	\begin{minipage}[b]{0.5\linewidth}
	\centering
	\epsfig{file=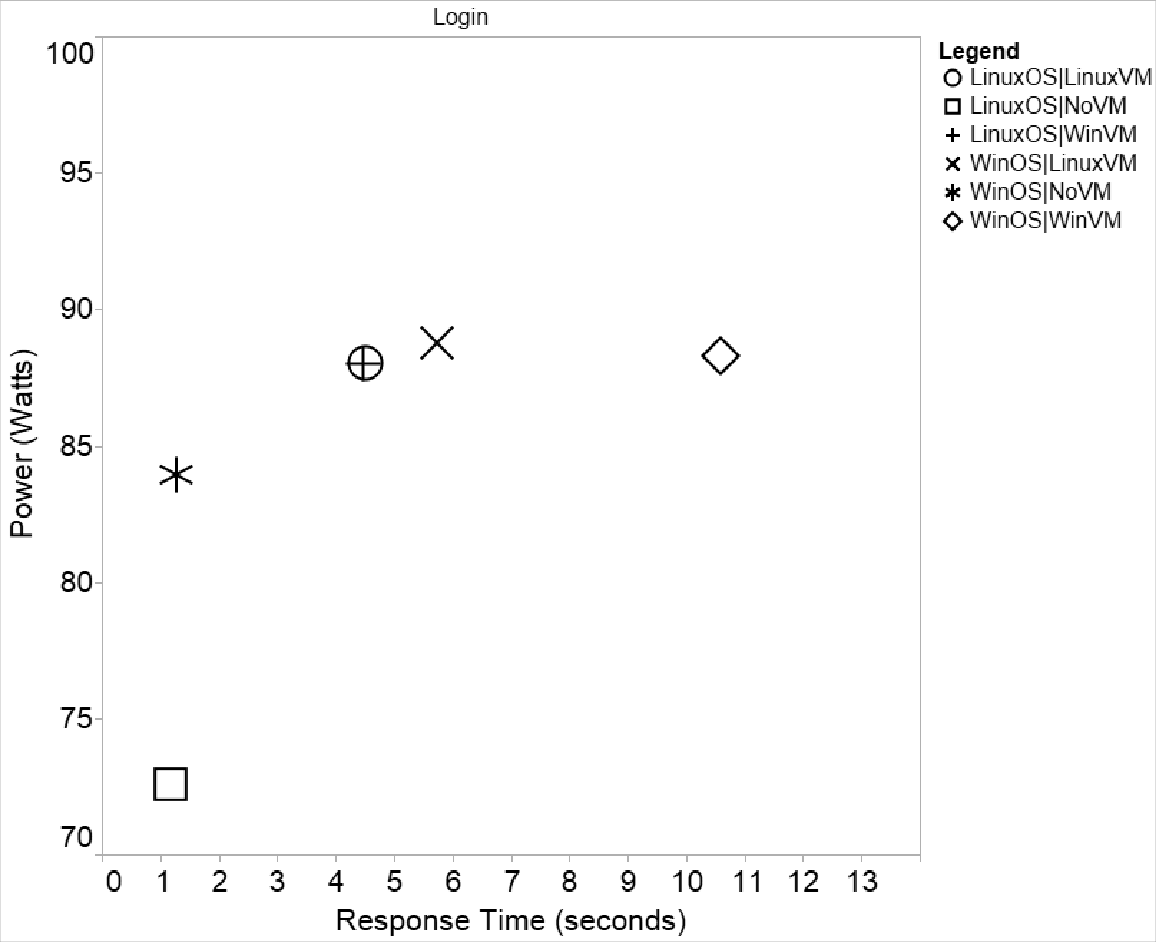,width=2.7in}
	\caption{Response time vs. Energy characteristics for Login actions}
	\label{fig:Login_RT_Energy}
	\end{minipage}
\end{figure*}

\begin{figure*}
	\centering
	\epsfig{file=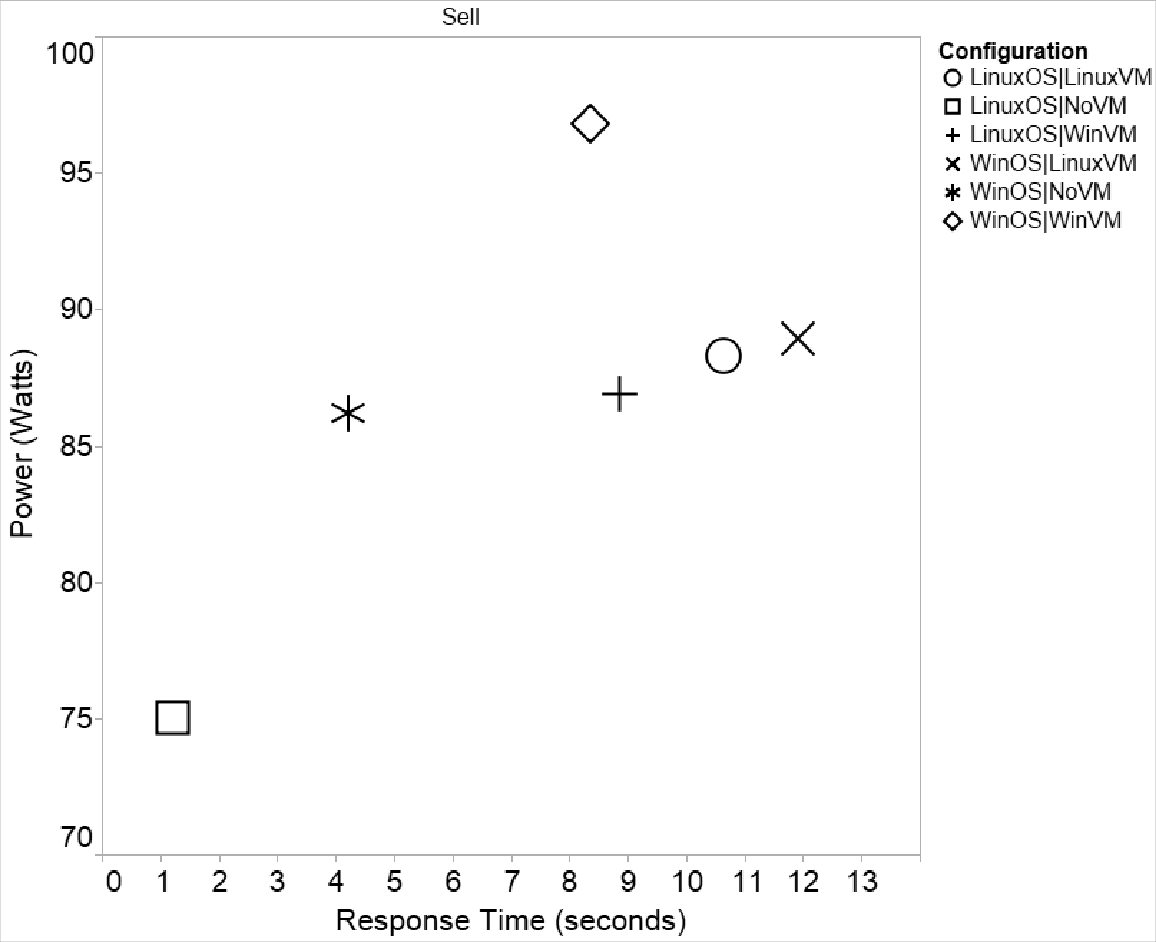,width=2.7in}
	\caption{Response time vs. Energy characteristics for Sell actions}
	\label{fig:Sell_RT_Energy}
\end{figure*}

In this section, we present the results of our experiments.  We consider each experimental case, i.e., browsing, logging in, and selling, in turn.  We plot energy usage (Figure~\ref{fig:browse-power}) and response time (Figure~\ref{fig:browse-rt}) for the browse action in the RUBiS application for increasing workloads, as well as a relative comparison of response time and power consumption for each configuration case for 1,000 concurrent JMeter users in Figure~\ref{fig:Browse_RT_Energy}.  Figure~\ref{fig:login-power} demonstrates the energy usage characteristics of the login scenario as workload increases.  Figure~\ref{fig:login-rt} depicts application response time for login with increasing workload.  Figure~\ref{fig:Login_RT_Energy} summarizes the response time and energy usage of the login action across all O/S configurations at 1,000 simultaneous JMeter users.  For the sell action, we plot energy consumption in Figure~\ref{fig:Sell-power}, and response time performance in Figure~\ref{fig:Sell-rt} for increasing workloads. We summarize the performance and energy usage across all experimental cases at 1,000 simultaneous users in Figure~\ref{fig:Sell_RT_Energy}.

We first consider the broad trends shown across all RUBiS actions.  For all actions, for the L and W cases provide both the lowest energy usage as well as the fastest response time results.  For all four cases where a VM and hosted O/S are configured, both response time and energy usage are higher than in the native O/S cases -- 9\% energy consumption and 22\%  response time on average.  This makes sense intuitively; adding two additional layers of a VM and hosted O/S to the stack should add overhead.  Without any information on how much energy overhead is involved, an architect may configure a standard VM-based architecture across all servers in an organization without understanding the consequences in terms of energy usage.  In such a scenario, the organization would incur a continuous energy penalty across all servers configured with a less energy-efficient platform.  

We next look at the impact of various OS and VM configurations for the RUBiS browse action for increasing workloads. The Linux native case (L) shows both the lowest energy usage as well as the fastest response time across all cases.  Let us next consider the L and W cases next.  Here, the response time performance of the application is virtually identical for both cases.  Absent any other performance indicator, an architect may decide to select W as a native OS platform. However, this would result in a significant increase in energy consumption, up to 16\% to 20\% greater for Windows compared to Linux, which would have a broad long-term impact on TCO (total cost of ownership) of the application for energy costs.  

We next consider the login action case.  The Linux native case provides the best performance in terms of both response time and energy consumption. The results show an interesting pattern in the cases of WW and LW. The LW case consumes slightly more energy than the WW case; however, the application's login response time is substantially higher in WW platform than in LW platform. Clearly, the response time is not proportional to energy consumption here.  Further, as in the browse action scenario, the W case shows the same energy consumption as the L case, but the W case consumes about 16\% more energy than the L case. This again demonstrates that considering only the response time performance of an application is not sufficient to judge the cost of the application in terms of power consumption; energy use should be considered independently.

Finally, we focus on the RUBiS sell action. For this action, the case of interest is WW, which provides a lower response times compared to the LL and LW cases, and the same response time as the WL case.  However, the WW scenario shows significantly higher energy consumption than all other experimental cases. Thus, an architect may select WW considering that it shows the best response time performance across all all VM architecture configurations for Sell actions.  However, such a decision would result in up to 12\% greater energy usage than other VM-based configurations. 

To summarize, our results here demonstrate that software architecture decisions, exemplified by a common choice of OS and VM configuration, can have a non-trivial impact on energy consumption, up to 10-20\% in our experiments here.

\section{Potential Impacts of Energy-Conscious Decisions}
\label{discussion}
In our experiments, we have shown that common, everyday software developer and enterprise architect choices can result in significantly different runtime energy consumption levels.  These are not one-time costs, but rather costs that are continually accrued over time over the production lifetime of an application. For instance, every time an application uses a more energy-expensive data type than is necessary, there is an energy penalty incurred -- potentially millions of times a day.  When an architect selects an off-the-shelf application that is more power-hungry than an equivalent option that consumes less power, the difference in energy efficiency results in a higher watt-hour consumption for every hour of the application's useful life in production.  Consider the average release cycle for a new version of the Windows operating system.  Historically, Microsoft releases a major new version of Windows every 3.4 years~\cite{windowsHistory:13}, which is a long time to continually incur a penalty for an energy-inefficient choice.

Although our experiments focus on how software impacts the power consumption on servers, the issue of energy usage is not limited to the servers themselves.  Indeed, there are significant infrastructure costs associated with supporting servers, including cooling and power dissipation losses~\cite{stanley:07}.  Data center managers use a heuristic, Site Infrastructure Energy Overhead Multiplier (SI-EOM), to estimate the infrastructure power usage associated with IT power usage.  This heuristic is a multiplier describing the additional power consumed by infrastructure support components for every unit of energy consumption by IT components, and is estimated to average 2.0.  That is, for every kWh drawn by a server, data center managers estimate an additional kWh of energy use by other components, e.g., cooling and other infrastructure~\cite{k:08}. 

If we consider the results of our experiments in the context of the SI-EOM, then every percentage point of energy savings a developer or architect can produce through more energy-conscious decision-making has a multiplicative effect through the supporting infrastructure.  Thus, if we in the IS community could reduce overall software-driven energy consumption in data centers by an average of 5\% through the options we choose, the multiplier effect would provide an additional 5\% energy savings in the infrastructure.  The effect would be a 10\% reduction in power consumption.  

We consider the potential impact of such a reduction in power consumption, taking the data center case as a case study, since energy usage here is well-documented.  Data centers are  estimated to account for 1\% to 1.5\% of worldwide energy consumption~\cite{k:08}.  A 10\% reduction would reduce these estimates to 0.9\% to 1.35\% of worldwide energy consumption.  A reduction of 0.1\% to 0.15\% may not seem like much at first glance, but it is actually quite significant.  A straightforward conversion~\cite{epa_conversion:12} to carbon emission shows that this would be equivalent to taking 4.5 million cars off the road.  

This opens the door to macro-level opportunities, such a carbon trading (e.g., as defined under the Kyoto Protocol or the European Union Emissions Trading Scheme), whereby entities that operate with fewer emissions than expected can ``trade" the right to emit greenhouse to less environmentally-friendly entities~\cite{c:12}.  

Further, on a more micro level, organizations have the opportunity to cite significant energy-efficiency gains on two of the three pillars of the triple bottom line (profit, people, planet)~\cite{econ:09}.  Specifically, organizations can cite the energy cost savings directly in the ``profit" bottom line, and the unused power (and related unspent emissions) in the ``planet" bottom line.  In a similar vein, organizations can cite such energy usage reductions as part of Corporate Social Responsibility efforts~\cite{ms:2001}.

How can we as members of the IS community achieve these types of energy efficiencies, as Watson et al.~\cite{wbc:10} and Murugesan~\cite{m:08} call us to do?  As we have shown in this work, there are energy impacts associated with very common IS/IT decisions.  It would seem likely that there are energy consequences associated with most, if not all, decisions made across the software lifecycle, including design, development, testing, selection, deployment, and retirement.  Currently, these energy consequences are not available to these decision-makers at the time that choices are made.  Clearly, there is a need for further research in this area to develop mechanisms, both technical and organizational, to make the energy usage ramifications of software choices available to the relevant decision makers throughout the software lifecycle.

\section{Conclusion}
\label{conclusion}
In this paper, we consider the energy efficiency of software applications, which has received little attention in the literature to date.  Specifically, we consider a set of common software developer and enterprise architect decisions, and describe the potential energy impacts across a range of options for each decision.  We demonstrate experimentally that there is significant potential for improved energy efficiency in software decision-making, both at design time and selection time.  If applied broadly, energy-conscious decisions can result in substantial savings in worldwide power consumption, with associated impacts in environmental benefits, as well as cost savings to organizations.  Thus, we argue for the need to consider energy consumption as a part of decision-making at every stage of the software lifecycle. 

\end{spacing}

\begin{small}
\begin{singlespacing}
\bibliographystyle{IEEEtran}
\bibliography{power}
\end{singlespacing}
\end{small}

\end{document}